\pdfoutput=1

\documentclass[fleqn,usenatbib]{mnras}

\usepackage[T1]{fontenc}

\DeclareRobustCommand{\VAN}[3]{#2}
\let\VANthebibliography\thebibliography
\def\thebibliography{\DeclareRobustCommand{\VAN}[3]{##3}\VANthebibliography}

\usepackage{graphicx}	
\usepackage{amsmath}	

\newcommand{\Fig}{{Figure}}

\graphicspath{figs/}

\title[Model of failed eruption]{A model of failed solar eruption
  initiated and destructed by magnetic reconnection}

\author[Jiang et al.]{Chaowei Jiang,$^{1,2}$\thanks{E-mail:
    chaowei@hit.edu.cn (CWJ)} Aiying Duan,$^{3}$ Peng Zou,$^{1}$ Zhenjun Zhou,$^{3}$ Xinkai Bian,$^{1}$ Xueshang Feng,$^{1,2}$
    \newauthor Pingbing Zuo,$^{1,2}$ and Yi Wang$^{1,2}$
  \\
  $^{1}$Shenzhen Key Laboratory of Numerical Prediction for Space
  Storm, Institute of Space Science and Applied Technology,\\ Harbin
  Institute of Technology, Shenzhen 518055, China\\
  $^{2}$Key Laboratory of Solar Activity and Space Weather, National
  Space Science Center, \\Chinese Academy of Sciences, Beijing 100190,
  China\\
  $^{3}$School of Atmospheric Sciences, Sun Yat-sen University, Zhuhai
  519000, China }

\date{Accepted XXX. Received YYY; in original form ZZZ}

\pubyear{2023}

\begin{document}
\label{firstpage}
\pagerange{\pageref{firstpage}--\pageref{lastpage}}
\maketitle

\begin{abstract}
  Solar eruptions are explosive disruption of coronal magnetic fields, and often launch coronal mass ejections into the interplanetary space. Intriguingly, many solar eruptions fail to escape from the Sun, and the prevailing theory for such failed eruption is based on ideal MHD instabilities of magnetic flux rope (MFR); that is, a MFR runs into kink instability and erupts but cannot reach the height for torus instability. Here, based on numerical MHD simulation, we present a new model of failed eruption in which magnetic reconnection plays a leading role in the initiation and failure of the eruption. Initially, a core bipolar potential field is embedded in a background bipolar field, and by applying shearing and converging motions to the core field, a current sheet is formed within the core field. Then, tether-cutting reconnection is triggered at the current sheet, first slow for a while and becoming fast, driving an erupting MFR. Eventually, the rise of MFR is halted by the downward magnetic tension force of the overlying field, although the MFR apex has well exceeded the critical height of torus instability. More importantly, during the rise of the MFR, it experiences a significant rotation around the vertical axis (with a direction contrary to that predicted by kink instability), rendering the field direction at the rope apex almost inverse to the overlying field. As a result, a strong current sheet is formed between the MFR and the overlying flux, and reconnection occurring in this current sheet ruins completely the MFR.
\end{abstract}

\begin{keywords}
  Sun: Magnetic fields -- Sun: Flares -- Sun: corona -- Sun: Coronal
  mass ejections -- magnetohydrodynamics (MHD) --  methods: numerical
\end{keywords}



\section{Introduction}
\label{sec:intro}

Solar eruptions, including mainly solar flares and coronal mass
ejections (CMEs) have been recognized as a leading driver of
disastrous pace weather in the solar-terrestrial space
environment. Now it is commonly believed that solar eruptions have
their root in the evolution of magnetic fields in the solar atmosphere,
in particular, the solar corona where magnetic fields dominate the
dynamics. Although not being directly observed, the coronal magnetic
fields in three-dimensional configuration have a high degree of
complexities as inferred from the plasma
emissions~\citep{aschwandenPhysicsSolarCorona2004}, and it remains
elusive how the solar eruptions are initiated. One of the complexities
is reflected in an interesting fact that solar eruptions can be
classified into two distinct types, namely eruptive flares and
confined flares. In eruptive flares, the eruption structures (plasma
and magnetic flux) can successfully escape into the solar wind and
form the so-called interplanetary CMEs. The confined flares are those
without association of CME, and they appear to be further divided into
two types. In some confined flares, only a series of brightening of
coronal loops are observed without much change in their shape during
the flare, for example in the well known X-class confined flares in the
super active region (AR) NOAA 12192~\citep{sunsunxudongWHYGREATSOLAR2015, jiangHOWDIDMAJOR2016}. In
other confined flares, the impulsive phase is similar to the eruptive
flares in that there are evident fast acceleration of magnetic
structures, i.e., erupting filaments, to certain heights, but the
erupting structures are quickly decelerated by some reasons and fail
to escape as a CME, and these events are known as failed filament
eruptions~\citep{jiObservationsFailedEruption2003, zhouWhyTorusunstableSolar2019, filippovFailedEruptionsSolar2020}. Therefore to
understand these failed eruptions, one should know how the magnetic
field is first initiated to rise impulsively and then how it is
trapped.


A prevailing model for failed eruption is that a coronal magnetic flux
rope (MFR) runs into kink instability but does not reach the threshold height for torus instability~\citep{torokConfinedEjectiveEruptions2005, hassaninhelicalkinkinstability2016, hassaninModelHomologousConfined2022}. MFR refers to a
coherent group of magnetic field lines winding around a common axis,
and is believed to a fundamental magnetic configuration in solar
eruptions~\citep{chengOriginStructuresSolar2017,
  guooriginstructuressolar2017, liumagneticfluxropes2020,
  patsourakosdecodingpreeruptivemagnetic2020,
  duanStudyPreflareSolar2019}. The kink instability occurs when the
twist degree of field lines exceeds a certain threshold, and during
its development, the MFR axis will be deformed with the increase of
its writhe at consumption of the magnetic field line
twist~\citep{hoodkinkinstabilitysolar1979,
  hoodcriticalconditionsmagnetic1981, torokidealkinkinstability2004,
  torokwrithehelicalstructures2010,
  torokConfinedEjectiveEruptions2005}. As a result, the MFR will rise
rapidly. It can erupt successfully if the apex of the axis reaches a
critical height above which the overlying field decays sufficently
fast with height. In such case, another instability due to the
interaction of the MFR and its strapping field, known as the torus
instability~\citep{kliemTorusInstability2006,
  demoulinCRITERIAFLUXROPE2010,
  aulanierFORMATIONTORUSUNSTABLEFLUX2010, faneruptioncoronalflux2010},
is triggered. Otherwise the erupting MFR will not be able to break
through the strong overlying field and thus results in failed
eruption. Both these two cases, i.e., the successful eruption and
failed eruption, have been shown
by~\citet{torokConfinedEjectiveEruptions2005} using MHD simulation
with initial condition given by a MFR model developed
in~\citet{1999A&A...351..707T}.

In this paper, we proposed a new model of failed eruption in which
magnetic reconnection plays the major role in the initiation and
failure of the eruption. This model is developed based on our recent
numerical simulations of a fundamental mechanism of eruption
initiation~\citep[][referred to as Paper A in the
following]{jiangFundamentalMechanismSolar2021} and also
in~\citep{bianHomologousCoronalMass2022,
  bianNumericalSimulationFundamental2022}, in which a successful
eruption can be triggered and driven by fast reconnection at a current
sheet that is formed in the quasi-static evolution of a
continually-sheared magnetic arcade. In this work, we show that the
eruption as initiated by the reconnection can be failed by adding an
overlying field with sufficient strength to the sheared arcade
(which is referred to as the core field). After the current sheet
is formed in the core field as driving by shearing and converging
motions, there is a short interval with slow reconnection at the
current sheet, through which a MFR is created. Then the reconnection becomes
fast, and the MFR rises impulsively, marking the onset of eruption. At the eruption onset, there is no evidence for either kink
instability or torus instability of the MFR, and the Lorentz force
analysis shows that the acceleration of the MFR is driven by the
slingshot effect of the reconnection. Eventually, the rise of MFR is
halted by the downward magnetic tension force of the background
overlying field, although the MFR apex has well exceeds the critical
height of torus instability. More importantly, during the rise of the
MFR, the interaction of the MFR with the background field renders a
fast rotation of the MFR around the vertical axis (with a direction
inverse to that as predicted by the kink instability) until its
direction at the apex of the rope is almost inverse to the overlying
field. As a result, a strong current sheet is formed between the MFR
and the overlying flux and reconnection occurs in this current sheet
ruins completely the MFR.

\begin{figure*}
  \centering
  \includegraphics[width=0.7\textwidth]{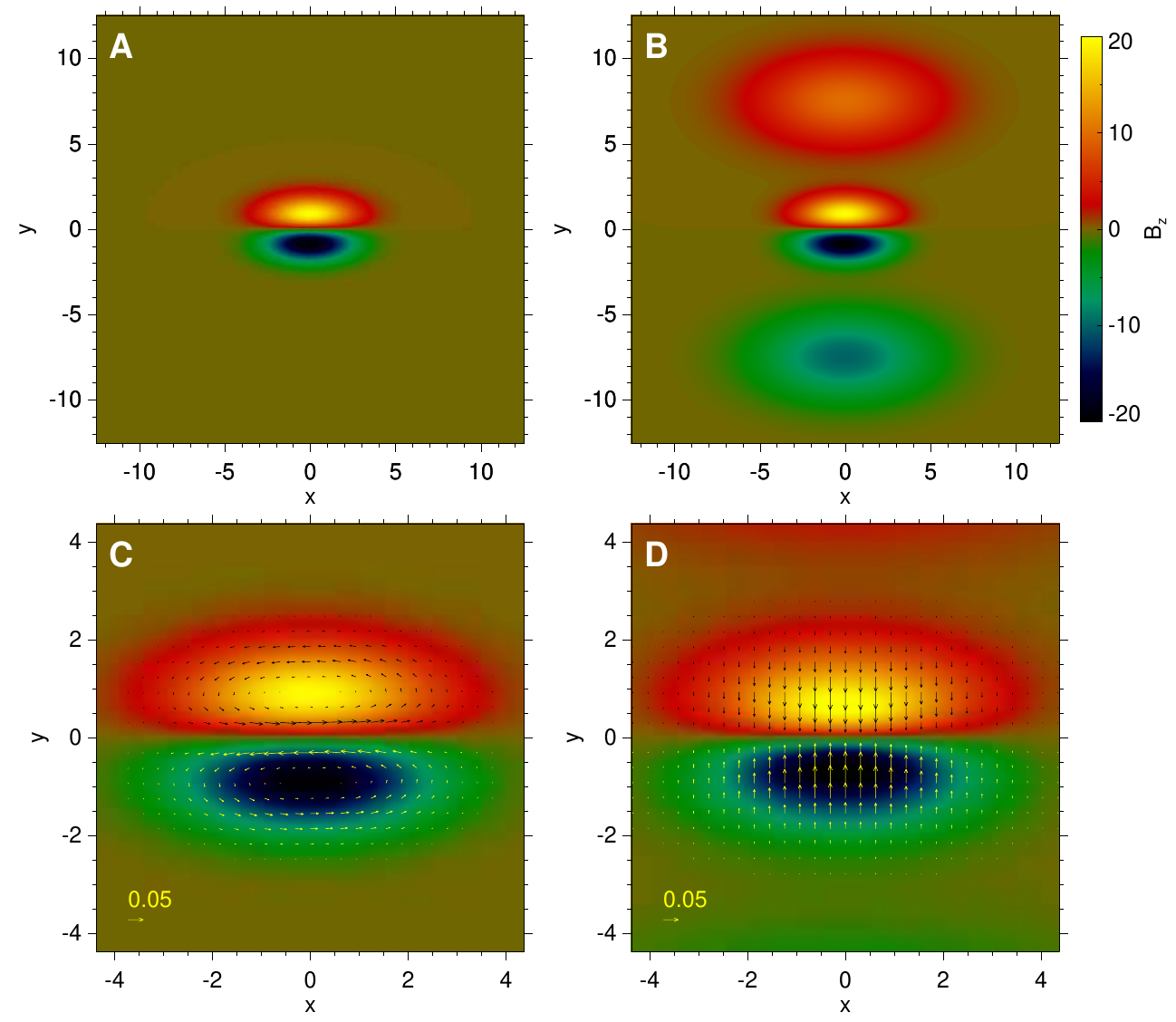}
  \caption{Magnetic flux distribution and driving flow at the bottom
    surface. (A) The case with only the core field. (B) The case with
    the core field and the background field. (C) Arrows show the
    surface rotational flow. (D) The surface converging flow.}
  \label{bottom_map}
\end{figure*}

\begin{figure*}
  \centering
  \includegraphics[width=0.7\textwidth]{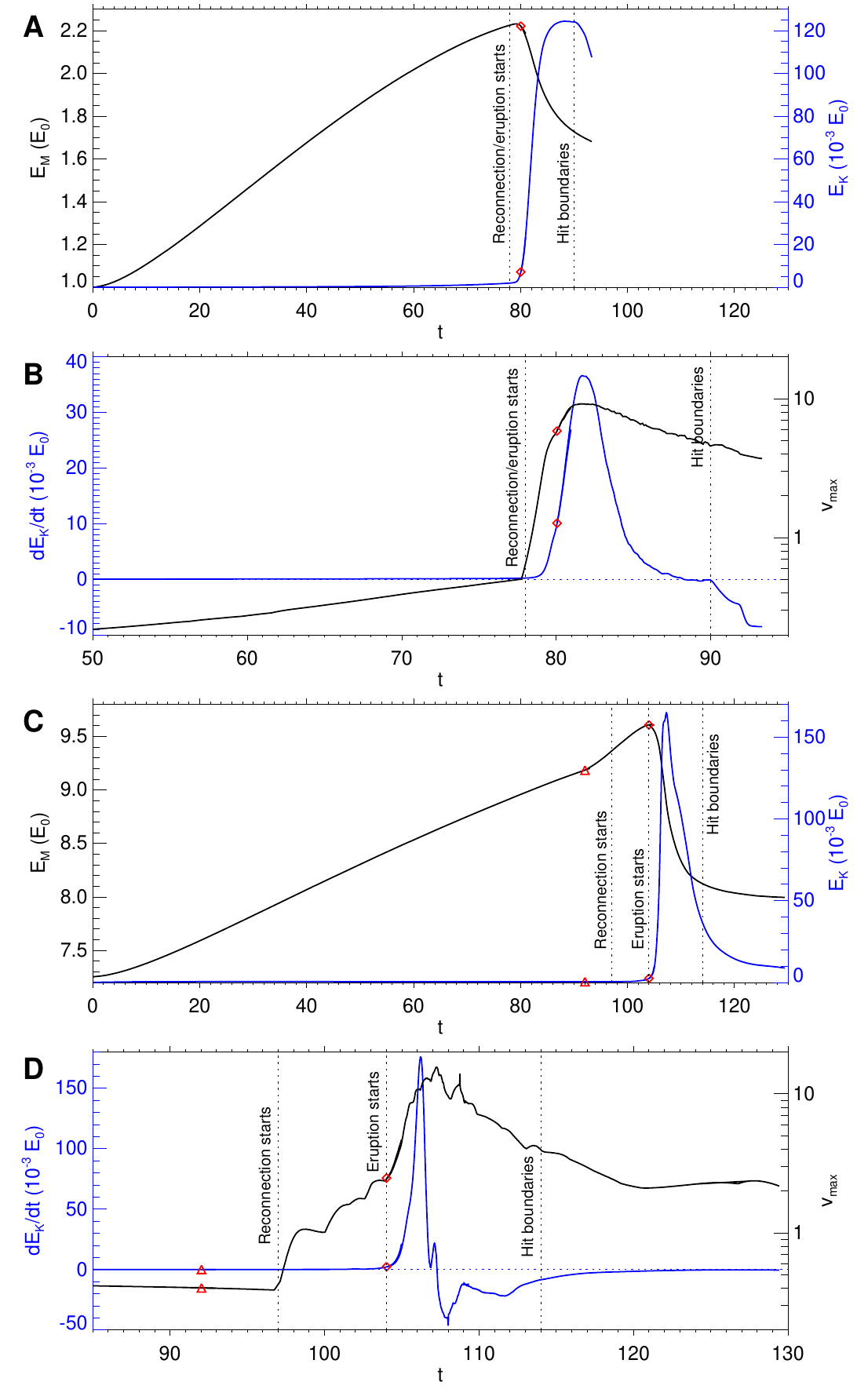}
  \caption{Temporal evolution of energies and velocity in the
    simulations. (A) Magnetic and kinetic energies in the case with
    only the core field. The energies are normalized by the potential
    field energy of the core field. (B) Time derivative of the kinetic
    energy and the maximal velocity in the case with only the core
    field. Note that the axis of the velocity is shown in logarithm scale
    for emphasizing its rapid changes with onset of reconnection. (C) and (D) have the same format with that of (A) and (B)
    respectively, but for the case with the background field. In the
    panels, the diamond symbol colored in red indicates that the
    surface driving is turned off. The triangle symbol in red
    indicates that the shearing motion is turned off and the
    converging motion is turned on. The critical stages are denoted by
    the vertical dashed lines, which include the start of
    reconnection, the onset of eruption and the time when the eruptive
    disturbances reach the lateral/top boundaries.}
  \label{paraevol}
\end{figure*}

\section{Method}
\label{sec:data}
The model is developed based on numerically solving the full set of
MHD equations with typical plasma and magnetic field settings for the
corona, which is essentially an environment with low plasma $\beta$,
very low resistivity and viscosity (corresponding to very high
Lundquist and Reynolds numbers), and high ratio of Alfv{\'e}nic to
sonic speeds. As the specific form of the equations has been given in
Paper A, we will not repeat it here. The variables are all
non-dimensionalized by typical values in the corona, which are
respectively, length $L_s = 11.52$~Mm, and time $t_s = 105$~s, density
$\rho_s = 2.29\times 10^{-15}$~g~cm$^{-3}$, temperature
$T_s = 10^6$~K, velocity $v_s = 110$~km~s$^{-1}$, magnetic field
$B_s = 1.86$~G. The computational volume spans a large Cartesian box
with both $x$ and $y$ extending $[-32, 32]$ and $z \in [0,64]$ (all
the values are expressed in non-dimensionalized numbers in this paper
if not mentioned particularly).  The full volume is resolved by a
block-structured adaptive mesh refinement (AMR) grid in which the
highest resolution is $\Delta =1/64$ (corresponding to $180$~km) to
capture the formation process of the current sheet and the
reconnection in the core field. The bottom surface of the
computational volume (the $z=0$ plane), which mimics the photosphere,
is assumed to be a line-tied boundary at which the footpoints of the
coronal magnetic field are anchored and convect with a prescribed
surface flow. At the bottom boundary, we directly solve the magnetic
induction equation to self-consistently update the magnetic field as
driven by the surface flow (which is implemented in the same way as
Paper~A). All other boundaries are open and nonreflecting by
zero-gradient extrapolation of variables from their neighboring inner
points. Same as Paper~A, we chose to not use explicit resistivity in
the magnetic induction equation, but magnetic reconnection can still
be triggered through numerical diffusion when a current layer is
sufficiently narrow (which supports a sufficiently strong current density) with thickness close to the grid resolution. By
this, we achieved an effective resistivity as small as we can with
given grid resolutions. Such setting also mimics the current-density-dependent resistivity as required for fast Petscheck-type reconnection.

At the beginning, a magnetic flux distribution at the bottom surface,
$B_z(x,y,0) = B_c + B_g$, is given by the superposition of a core
bipole $B_c$ and a background bipole $B_g$, both of which have the
same form defined as
\begin{equation}\label{eq:core_field}
		B_i = b_i e^{-x^2/\sigma_i^2}(e^{-(y-y_i)^2/\delta_i^2}-e^{-(y+y_i)^2/\delta_i^2}),
\end{equation}
where the subscript $i$ denotes $c$ or $g$. For the core bipole
$\sigma_c = 2$, $\delta_c = 1$, $y_c = 0.1$, and $b_c=117.4$ is a
scaling factor such that the maximum of $B_{c}$ is $20$, and for the
background bipole $b_c = 10$, $\sigma_c = 4$, $\delta_c = 2$,
$y_c = 6$.  The core field is a compact bipole, as shown in
\Fig~\ref{bottom_map}A, while the background field is a dispersed
bipole with much larger extents (\Fig~\ref{bottom_map}B). The ratio of
unsigned fluxes between the core field and the background field is
$0.33$. Then, a potential field is calculated for the bottom flux
distribution using the Green's function method. With this potential
field and a plasma in hydrostatic equilibrium as the initial
condition, we drive the MHD system to evolve using surface motion at
the lower boundary (i.e., driving motion). We have carried out two
simulations, one simulation (with both the core field and the
background field) that produced a failed eruption, and the other with
only the core field that can produce successful eruption. The
simulations consists of different phases, which are respectively the
shearing phase, the converging phase, and the non-driving phase, as
defined by different driving motion. In the shearing phase, we use a
rotational flow applied to each of the two magnetic polarities in the
core field. The velocity is defined by
\begin{equation}\label{eq:dirven_speed}
	(v_{x}, v_{y}) = (\dfrac{\partial \psi}{\partial y}, -\dfrac{\partial \psi}{\partial x});
     \psi = r_{0}B_{c}^{2}e^{-(B_{c}^{2}-B_{c, {\rm max}}^{2})/B_{c, {\rm max}}^{2}},
\end{equation}
where $B_{c,{\rm max}}$ is the largest value of the $B_{c}$ and
$r_{0}$ is a constant for scaling such that the maximum of the surface
velocity is $0.06$. This rotational flow will not modify the flux
distribution at the bottom surface. In the converging phase, we apply
a converging flow to the core field, defined as
\begin{equation}\label{eq:dirven_speed}
			(v_{x}, v_y) = (0,  - c_0 B_c)
\end{equation}
where $c_{0}$ is also a constant for scaling such that the maximum of
the surface velocity is $0.06$. This can drive the flux towards to the
PIL and help to build up a current sheet above the PIL. In the
non-driving phase, we stop all the driving motion and let the system
to evolve spontaneously. This can test whether the system has reached
a critical phase of eruption initiation.

\begin{figure*}
  \centering
  \includegraphics[width=\textwidth]{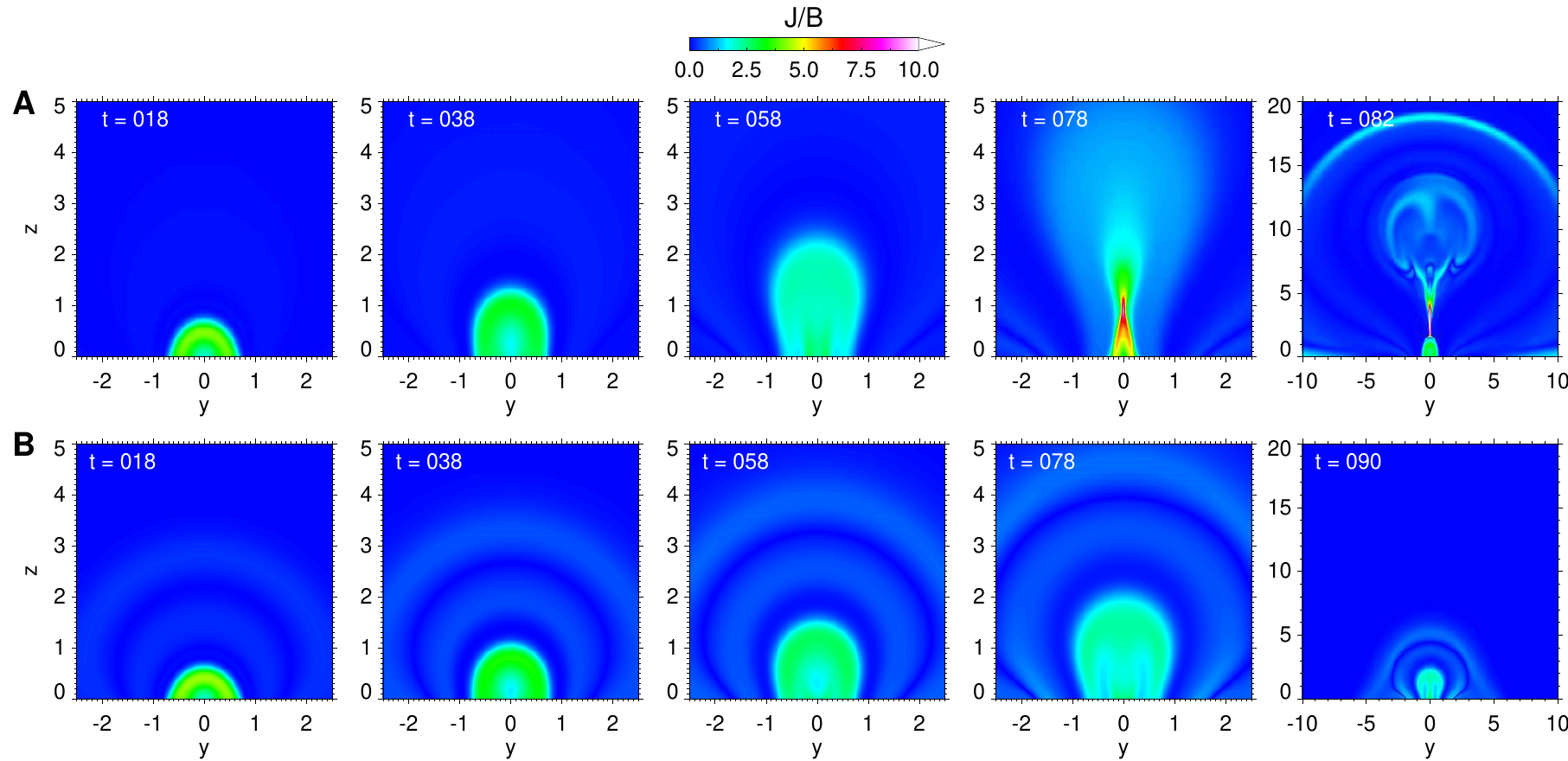}
  \caption{Comparison of the current distribution in the central
    vertical slice (i.e., the $x=0$ plane) for two simulation cases in
    the shearing driving phase. (A) The case with only the core field
    which produces a successful eruption. Note that at around $t=78$ a
    current sheet is formed, and then the eruption is immediately
    triggered by reconnection at the current sheet. (B) The case with
    the background field. No current sheet is formed even the
    shearing driving is applied until $t=90$.}
  \label{slice_compare}
\end{figure*}

\begin{figure*}
  \centering
  \includegraphics[width=\textwidth]{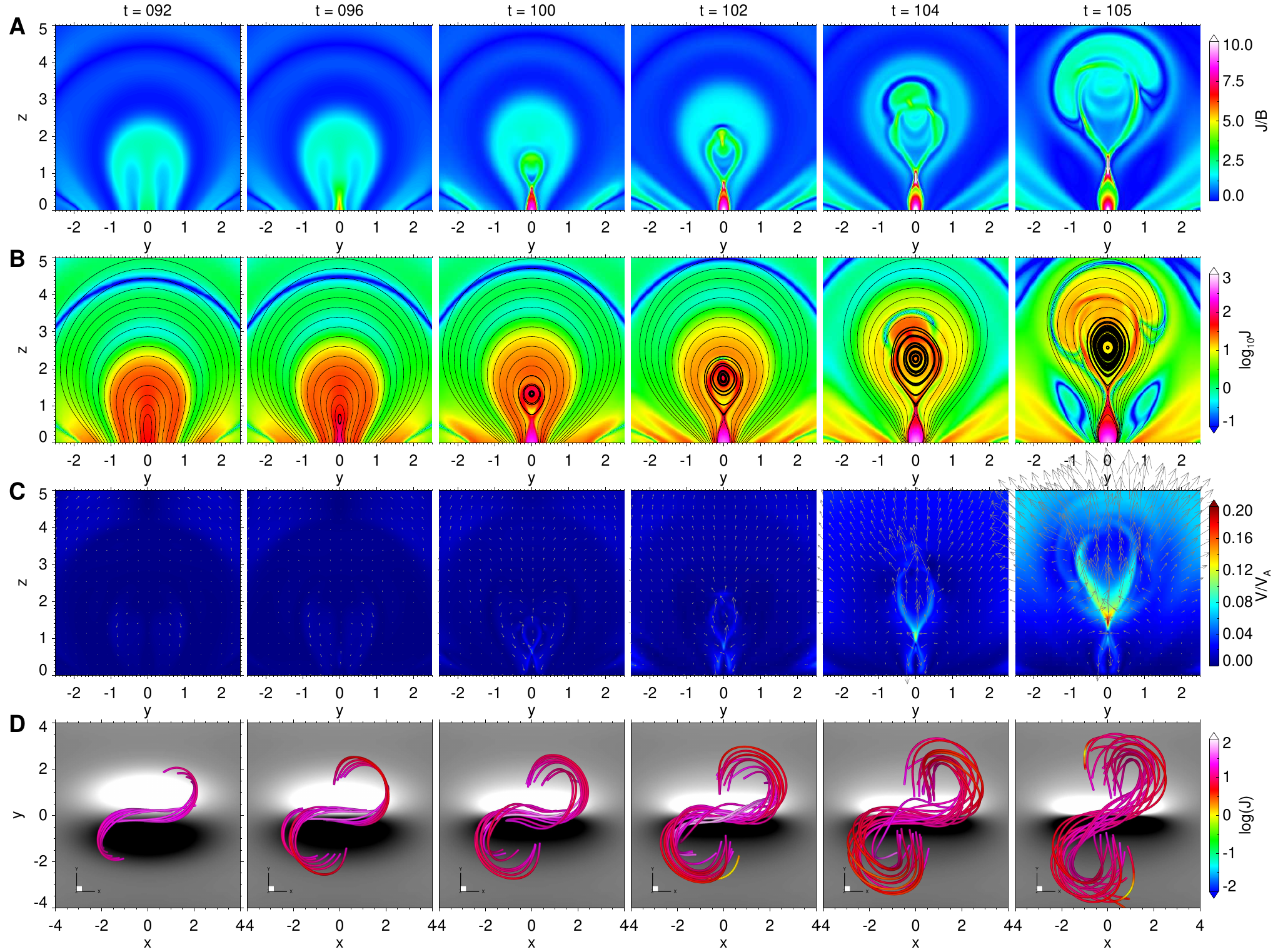}
  \caption{Formation of a pre-eruption MFR by slow tether-cutting
    reconnection until the eruption onset in the simulation of the
    failed eruption. (A)--(C) shows the current density normalized by
    magnetic field ($J/B$), the current density, and the Alfv{\'e}nic
    Mach number on the central vertical slice (the $x=0$ slice). In
    (B), the curves show the projection of the magnetic field lines on
    the slice. In (C), the arrows denote the velocity. (D) Sampled
    magnetic field lines showing the tranformation from the sheared
    arcade to the MFR. The color of the field lines indicates the
    current density. The background shows the magnetic flux
    distribution at the bottom boundary.}
  \label{mfr_form}
\end{figure*}

\begin{figure*}
  \centering
  \includegraphics[width=\textwidth]{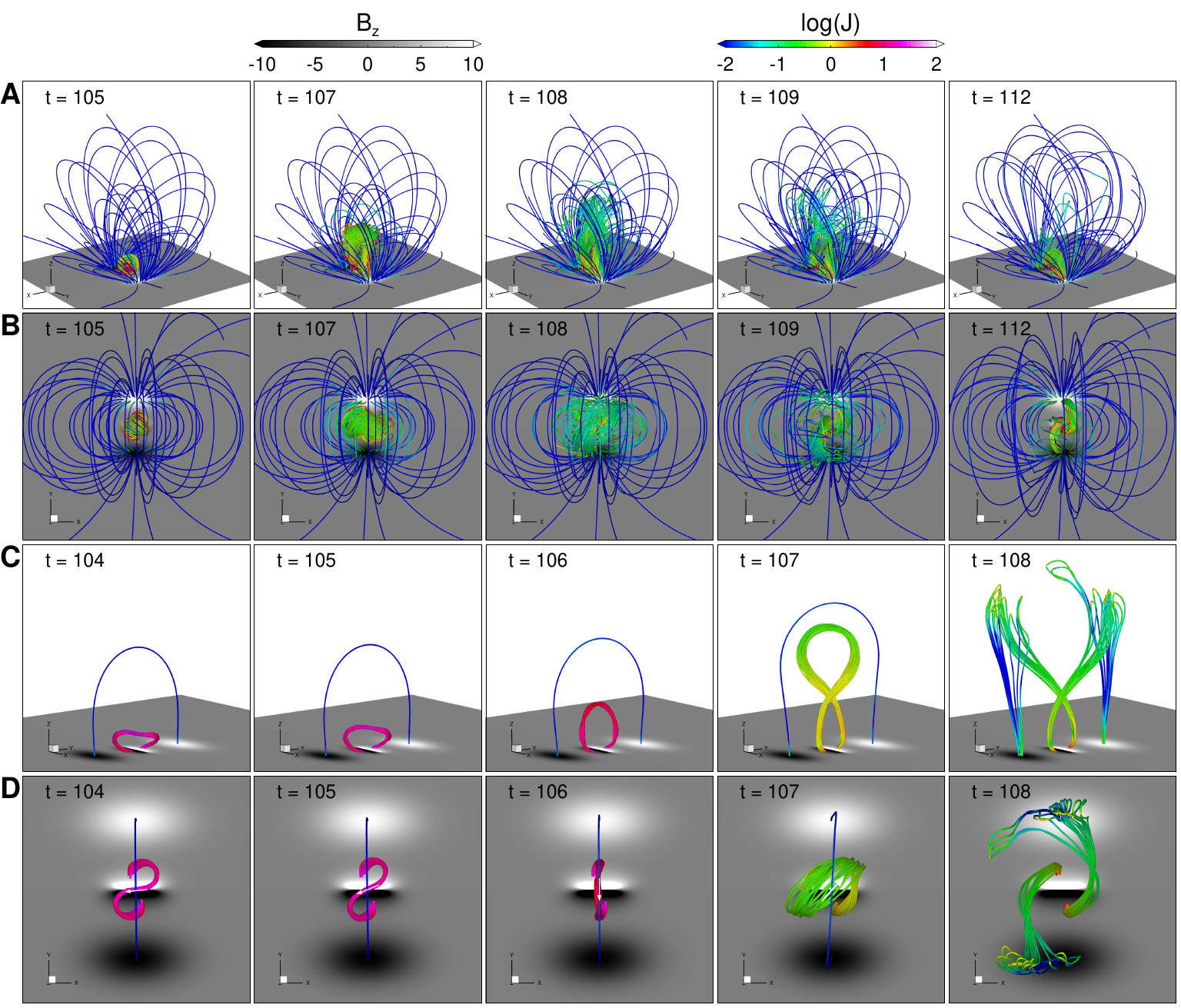}
  \caption{Evolution of magnetic field lines in the failed
    eruption. (A) 3D prospective view of field lines in the full
    region. The field lines are pseudo-colored by the current
    density. The bottom surface is shown with the magnetic flux
    distribution $B_z$. (B) Same as (A) but viewed from the top. (C)
    Close view of the erupting MFR. A field line of the background
    field is shown. (D) Same as (C) but viewed from the top. The
    arrows show the direction of the magnetic field lines of the
    MFR. All the field lines are shown with one of their footpoints
    fixed on the bottom surface, since no surface motion is applied
    during the eruption stage. An animation is available for this
    figure.}
  \label{eruption_3dflines}
\end{figure*}

\begin{figure*}
  \centering
  \includegraphics[width=0.7\textwidth]{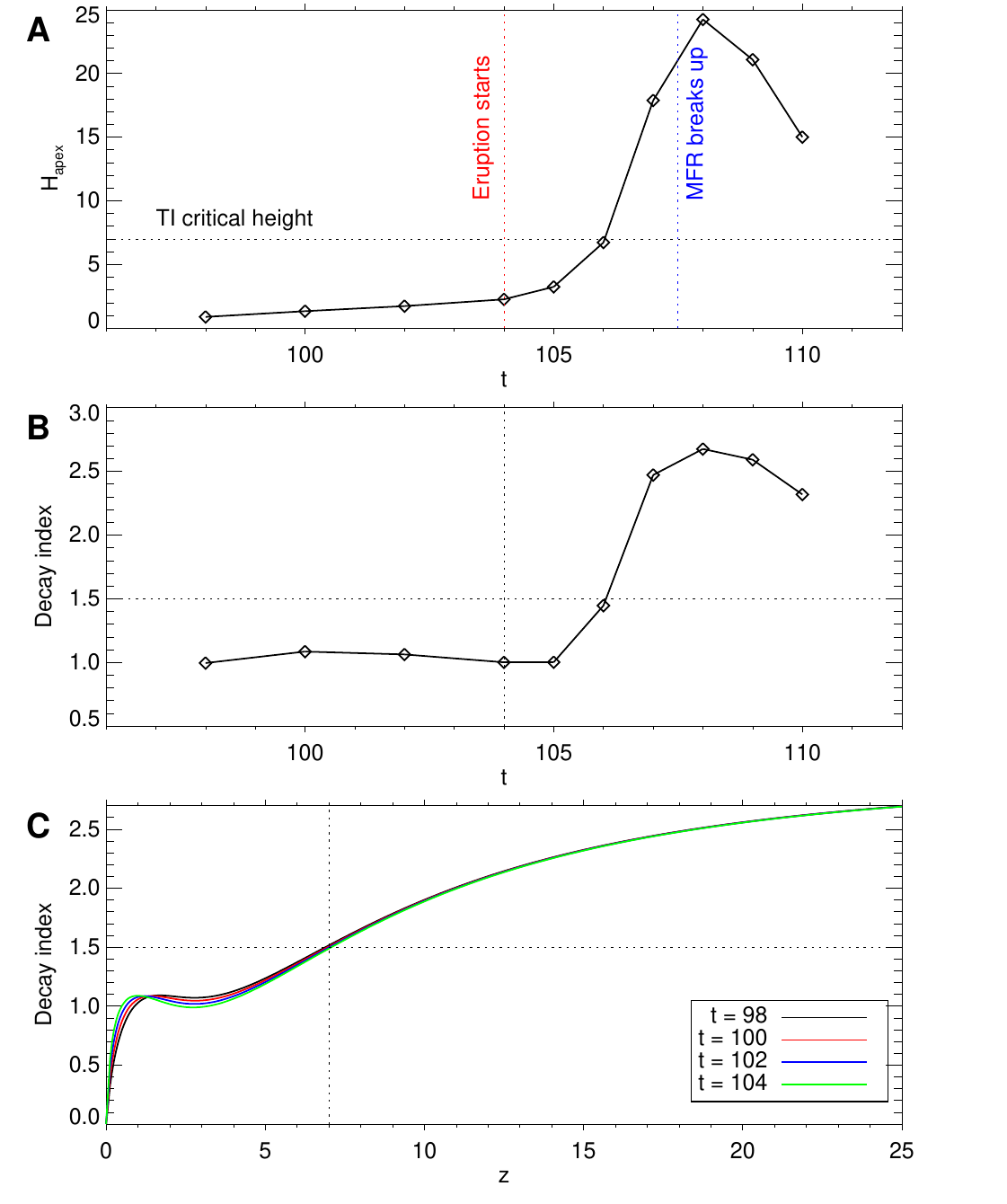}
  \caption{Evolution of the height and decay index of the MFR
    axis. (A) Temporal evolution of the apex height of the MFR
    axis. Since the MFR breaks up after $t=107$, the height is
    recorded for apex of the field line rooted in the footpoints of
    the original MFR axis. (B) The decay index of the potential field
    at the apex of the MFR axis. (C) The height profiles of the decay
    index during the evolution. Note that the converging motion at the
    bottom surface changes the magnetic flux distribution and thus the
    decay index also varies with time. The height and value of the
    critical decay index $n=1.5$ are shown by the dashed lines.}
  \label{MFRaxisheight}
\end{figure*}

\begin{figure*}
  \centering
  \includegraphics[width=\textwidth]{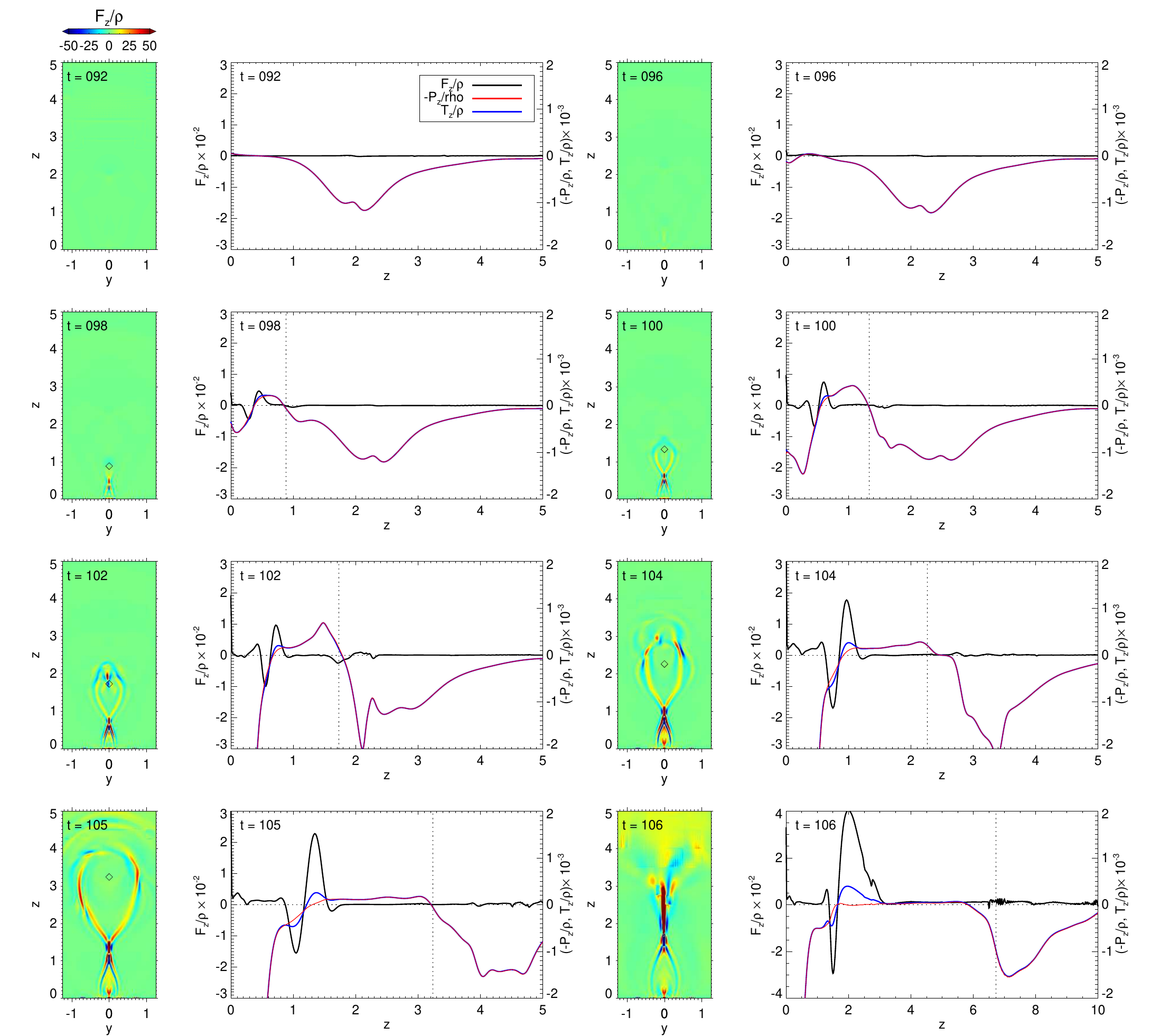}
  \caption{Distribution and evolution of the vertical Lorentz
    force. For each time, the central vertical slice is shown with the
    distribution of the vertical acceleration by the Lorentz force
    $F_z/\rho$, and the profiles along the central vertical line
    (i.e., the $z$ axis) are shown for the $F_z/\rho$ (the black
    line), the magnetic pressure-gradient force $P_z/\rho$ (the red
    line), and the magnetic tension force $T_z/\rho$ (the blue
    line). Note that $-P_z/\rho$ is plotted for a better comparison of
    its magnitude with that of $T_z/\rho$. Thus when the red and blue
    lines coincide, the two force components balance each other and
    $F_z/\rho$ is zero. The diamond symbols on the slices denote the
    location of the MFR axis, and its height is also shown by the
    vertical dashed lines in the line plots.}
  \label{lorentz_slice}
\end{figure*}

\begin{figure*}
  \centering
  \includegraphics[width=\textwidth]{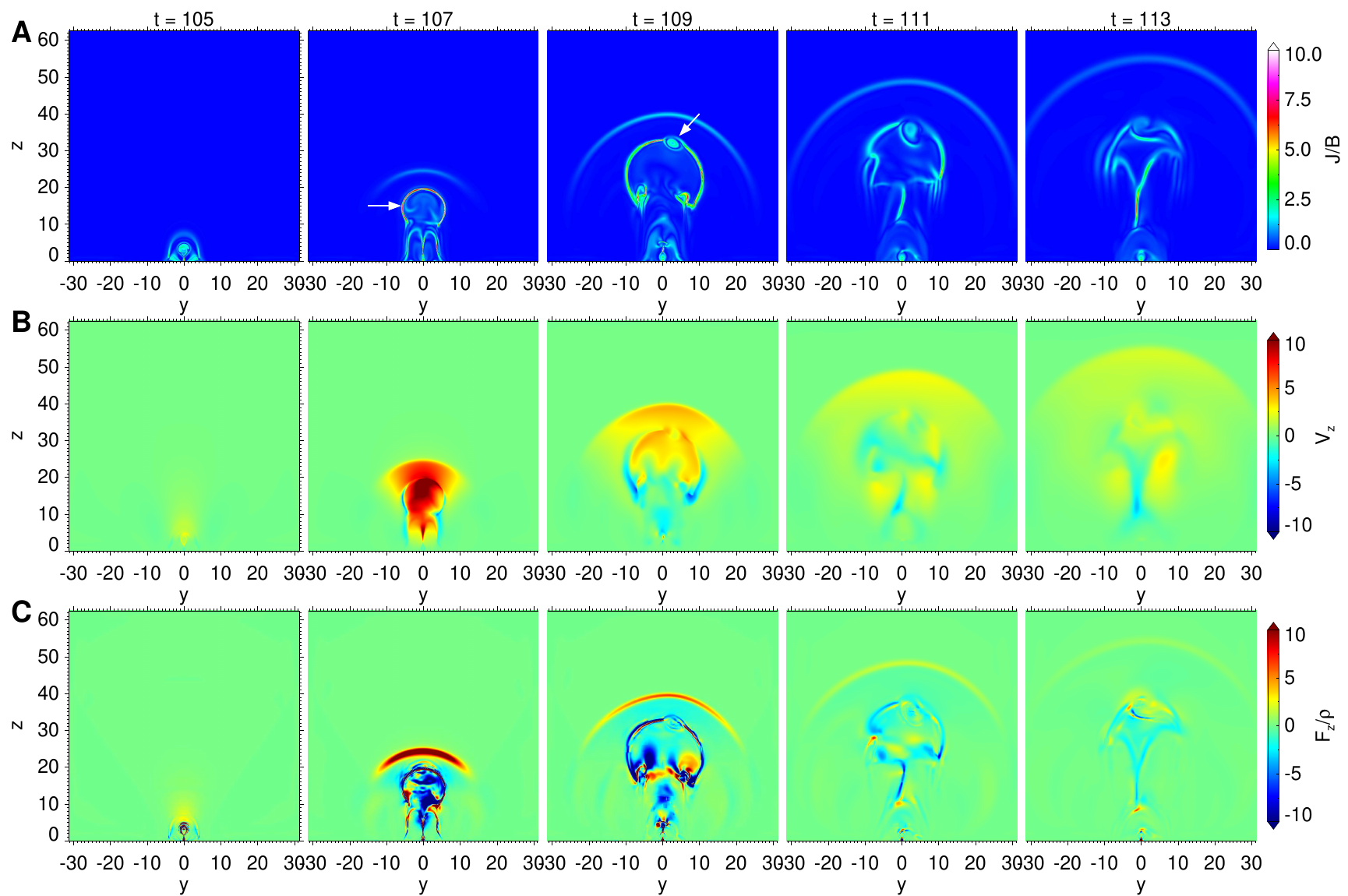}
  \caption{Evolution of different parameters on the central vertical
    slice of the failed eruption. (A) Current density normalized by
    the magnetic field strength. (B) Vertical component of the
    velocity. (C) Vertical component of the Lorentz force. The two arrows denote
    the external current sheet, and the plasmoid in the current sheet, respectively. An
    animation is available.}
  \label{eruption_slice}
\end{figure*}

\section{Results}
\label{sec:res}

\subsection{Evolution of energies}
\Fig~\ref{paraevol} shows temporal evolutions of the total magnetic
energy $E_M$,  the total kinetic energy $E_K$ and the maximal
velocity $v_{\rm max}$ in the different phases of the simulations.
The sharp variation of the maximal velocity with time can
indicate the onset of reconnection, since magnetic reconnection can impulsively
accelerate the plasma.
The
results are shown for both the simulation of only the core field
(\Fig~\ref{paraevol}A and B) and the one with the core field embedded
in the background field (\Fig~\ref{paraevol}C and D). The values of
all the energies are normalized by the potential field energy of the
core field (named as $E_0$). In both simulations, the evolution of
magnetic energy consists of two distinct stages, a slow accumulation
stage and a fast release stage. The transition point of the two stages
marks the onset of eruption, which is $t=78$ in the simulation with
only the core field and $t=104$ in the simulation with the background
field. In the pre-eruption stage, the MHD system evolves in a
quasi-static way, since the kinetic energy keeps a very low level of
mostly below $10^{-3}E_0$. While with the onset of the eruption the
kinetic energy impulsively increases, reaching a peak value of around
$10^{-1}E_0$. The behaviors after then are rather different between
the two cases. In the case with only the core field, the kinetic
energy keeps unchanged until the eruptive structure hits the side/top
boundaries of the computational volume, and then decreases because the
eruption leaves the volume freely. This shows that the eruption is
successful. In contrast, the kinetic energy in the case with the
background field rapidly declines, with a majority part (only $75\%$)
lost before any disturbance reaching the numerical boundaries, which
shows that the eruption is failed and trapped within the computational
volume.

\subsection{Simulation with no background field}
For the simulation with only the core field, its eruption is produced
by applying the continuous shearing at the bottom surface until the
eruption onset time. The eruption is initiated through the same
mechanism as demonstrated in a series of our previous
studies~\citep[Paper A, and][]{bianNumericalSimulationFundamental2022,
  bianHomologousCoronalMass2022,
  bianRolePhotosphericConverging2022}. That is, through
quasi-statically surface shearing motion along the PIL, a vertical
current sheet forms slowly above the PIL, and once the current sheet
is sufficiently thin such that ideal MHD is broken, reconnection sets
in and immediately triggers the eruption (see
\Fig~\ref{slice_compare}A, and also the detailed analysis of Paper
A). In Paper A, it is also confirmed that the reconnection results in
a strong upward tension force which plays the key role in driving the
eruption, that is, the slingshot effect of the reconnection is the
main driver of the eruption (see the last section in Methods of Paper
A). We note that this mechanism is different from the tether-cutting eruption initiation
model~\citep{mooreOnsetMagneticExplosion2001} in a twofold
sense. Firstly, the tether-cutting model is a two-step reconnection
process, in which it is assumed that before the onset of the eruption
(i.e., the start of impulsive phase of eruption), there is a phase of
slow reconnection from more than ten minutes to a few hours, which
gradually cuts the tethers, and it is not until a ``global
instability'' occurs that the eruption can start. Secondly, the
tether-cutting model proposed that the reconnection only plays the
role of cutting the confinement of the field lines and it is the
unleashed magnetic pressure that drives the eruption.

\subsection{Simulation with background field}
For the case with the background field, no eruption is resulted during
the shearing phase (which is applied until $t=92$), since the current
sheet has not yet formed, as shown in \Fig~\ref{slice_compare}B. The
reason is that, due to the constraining effect of the overlying
background field, the core field cannot strecth outward sufficiently
like the case without the background field (compare \Fig~\ref{slice_compare}B with A).
In principle, a current sheet can eventually form if a sufficiently
large amount of free energy is injected into the system, such that the
total magnetic energy is close to that of a fully open field, but this
might not be realized by only shearing the core field using the
rotational flow.  Then to create a current sheet more efficiently, we
apply the converging motion to the core field starting from
$t=92$. The converging flow injects magnetic energy into the corona
with a faster rate than the shearing motion
(\Fig~\ref{paraevol}C). \Fig~\ref{mfr_form} shows that, by the
converging flow, the core field is compressed towards to the PIL, and
consequently a current sheet forms at around $t=97$, with a lower height than that
in the case with only the core field.

Subsequently, reconnection is triggered at the current sheet, as shown
by the topology change of the core field from sheared arcade to flux
rope (\Fig~\ref{mfr_form}B) and the reconnection flows
(\Fig~\ref{mfr_form}C). The start of reconnection is also indicated by
the sudden rise of the maximal velocity as shown in \Fig~\ref{paraevol}D, at $t=97$.
But interestingly, unlike the case without the
background field, this reconnection does not immediately initiate an
eruption because the stabilizing effect of the overlying
flux. The reconnection occurs slowly as the Alfv{\'e}nic
Mach number of the reconnection outflow is mostly smaller than $0.1$
before $t=104$. This actually is line with the first step reconnection
of the tether-cutting model, i.e., the slow reconnection phase, which
creates a stable MFR. The slow reconnection takes a short
interval from $t=97$ to $104$, and during this interval, the maximal velocity $v_{\max}$ shows also
an intermittent increase and decrease, maintaining within a low level of up to $2$ (\Fig~\ref{paraevol}D), which reflects the fact that the
reconnection is confined by the overlying field. This is clearly different from the case with only the core field, in which the reconnection becomes fast once starts as the $v_{\max}$ impulsively rises to Alfv{\'e}nic speed of close to $10$ (\Fig~\ref{paraevol}C).
After the stage of slow reconnection, an eruption starts with continual and fast rise of the $v_{\max}$ to also an Alfv{\'e}nic value.
 \Fig~\ref{eruption_3dflines} shows
the eruption of the MFR and its final breaking up during its
interaction with the background field.

\subsection{Ideal instabilities of the pre-eruption MFR}
To check whether the eruption is triggered by the torus instability of
the MFR, in \Fig~\ref{MFRaxisheight} we traced the apex height of the axis of the MFR and calculated the decay index of the potential field (which
approximates the strapping field of the MFR). The decay index is defined as
$n(z)=-\partial (\ln B)/\partial \ln z$ along the $z$ axis. As can be
seen in \Fig~\ref{MFRaxisheight}A, the MFR experiences a slow rise
from around $z=1$ to $2$ before the eruption, as driven by the slow
reconnection. Once the eruption begins, it rises rapidly to a peak
height of around $z=20$ and then breaks up (due to reconnection with
the background field, to be analyzed below). After the MFR breaks up,
the height is recorded for apex of the field line rooted in the
footpoints of the original MFR axis. \Fig~\ref{MFRaxisheight}B shows
that from the pre-eruption to the early phase of the eruption
(until $t=105$), the decay index at the apex of the MFR axis is
$n=1\sim 1.1$. This value is lower than the canonical threshold ($1.5$) of
torus instability, and also below the typical unstable range of
$1.5 \pm 0.2$ as derived in many theoretical and numerical
studies~\citep[e.g.,][]{kliemTorusInstability2006,
  fanOnsetCoronalMass2007, aulanierFORMATIONTORUSUNSTABLEFLUX2010,
  zuccarelloCRITICALDECAYINDEX2015} as well as statistical
observational studies~\citep{duanStudyPreflareSolar2019,
  duanVariationMagneticFlux2021}, which suggests that the eruption is
not initiated by torus instability. Note that from $t=100$
to $104$, the decay index shows a mild decrease although the MFR
rises. This is because the height profile of the decay index has a
saddle-like shape~\citep{luoWhereHowDoes2022} with the transition
point at around $n=1\sim 1.1$, as can be seen in
\Fig~\ref{MFRaxisheight}C.

When seen from above (\Fig~\ref{eruption_3dflines}D), the MFR axis
initially has an inverse S shape with its middle part aligned with
the PIL. While during the eruption, it undergoes a fast
counterclockwise rotation. For instance, at $t=106$, which is the peak
time of acceleration of the eruption (see \Fig~\ref{paraevol}D), it
has rotated by nearly $90^{\circ}$ and is thus straightened overall,
becoming perpendicular to the PIL. The transformation of an erupting
MFR from a reversed S to a straight shape indicates a decrease of the
writhe of the MFR, which excludes the kink instability of the
MFR. Otherwise, the writhe should increase since the kink instability
converts magnetic twist into the writhe of the MFR axis. The
counterclockwise rotation of the erupting MFR with initially a reverse
S shape also occurs in the successful eruption without the background
field, which has been carefully studied by~\citet{Zhou_2022} based on the simulation in Paper A. They found that this rotation is
consistent with the rotation of a typical filament eruption observed
from dual viewing angles. Furthermore, by analyzing the Lorentz force
and torque, \citet{zhouMechanismMagneticFlux2023} confirm that the
external shear-field component rather than the internal force of the
MFR plays a key role in driving the counterclockwise rotation.

\subsection{Driver of the eruption}
Here, same as the finding in Paper A, the eruption is initiated by
reconnection, and in particular, driven by the strong upward tension
force of the newly reconnected field lines. To illustrate this, in
\Fig~\ref{lorentz_slice} we show the upward acceleration by the
Lorentz force $F_z/\rho$ on the central vertical cross section of the
volume and the profiles of the two force components (the magnetic
pressure-gradient force $P_z/\rho$ and tension force $T_z/\rho$) on
the central vertical line from pre-eruption to the peak time of
acceleration of the eruption. The slices show that the upward acceleration by the Lorentz
force is mainly distributed on the edge of the expanding MFR, i.e., in
the newly reconnected flux that joins in the MFR, while the
acceleration at the core of MFR is very weak ($F_z/\rho$ is almost
zero). In particular, the strongest upward
acceleration always resides in the reconnecting current sheet and acts
as the central engine of eruption. By decomposing the force into
magnetic tension force and pressure-gradient force on the central
vertical line, one can see that it is always the magnetic tension
force at the reconnecting outflow region of the current sheet (i.e.,
the slingshot effect of reconnection) that plays the role of upward
acceleration. Since only in this region, the tension force (the blue
curve) is not balanced with the pressure-gradient force (the red
curve), while elsewhere they almost cancel out with each other. Such
unbalance is directly caused by the reconnection which starts at
$t=97$, since before that the forces are perfectly balanced everywhere,
see $t=92$ and $96$ in \Fig~\ref{lorentz_slice}. In the early stage of
the reconnection, the force unbalance is rather weak (see $t=98$,
$100$, and $102$) and is not able to drive a fast eruption, because
the newly reconnected field lines will experience a quick deceleration
by the downward magnetic pressure-gradient force in the region between
the top of the reconnection outflow and the MFR axis. Therefore, the
slow reconnection proceeds in this stage, until the force unbalance is
sufficiently large to result in a disruption with a fast
reconnection. At the peak time ($t=106$), the acceleration of the slingshot effect is extremely rapid,
reaching up to $a_z \approx 200$ (which corresponds to
$\sim 200$~km~s$^{-2}$) within a height range of $L_z \approx
1$. Thus, as the field line approaches the current sheet and
reconnects, its middle point can be accelerated upward to a velocity
of around $\sqrt{2a_z L_z} = 20$ (which is consistent with the largest
speed as shown in \Fig~\ref{paraevol}D).

\subsection{Cause of failure of the eruption}
From the time profile of the decay index (\Fig~\ref{MFRaxisheight}B), it seems that the MFR enters
into the domain of torus instability after $t=106$ (during the eruption)
and reaches a peak height of $\sim 20$ with decay index of
$\sim 2.5$, far above the torus instability threshold. Nevertheless,
it does not erupt successfully owing two reasons. Firstly, the
erupting flux rope is decelerated by the downward magnetic tension
force of the background field overlying the core
field. \Fig~\ref{eruption_3dflines}C and \Fig~\ref{eruption_slice} show that when the erupting MFR
reaches the height of $\sim 20$, the background field is strongly
pushed upward, which in turn results in downward tension force, and
consequently the velocity decreases quickly. Secondly, as the MFR
rotates about the $z$ axis significantly during its eruption, the
field lines at its apex are even reversed in direction to that of the overlying
field (\Fig~\ref{eruption_3dflines}D). As a result, a strong current sheet is formed between the MFR
and the overlying flux (see $t=107$ in \Fig~\ref{eruption_slice}A), and reconnection
occurs in the current sheet, which destructs the MFR eventually into
two arcades connecting both the core and the background polarities
(see \Fig~\ref{eruption_3dflines}C and D).  Interestingly, as this
current sheet is rather large in scale, a plasmoid instability is
triggered and thus a new small flux rope is formed in the current
sheet during the reconnection (see the panel of $t=109$ in
\Fig~\ref{eruption_slice}A).

\section{Discussions}
We have presented a model for failed solar eruption based on numerical
MHD simulations of an eruptive core field embedded in a large-scale
background field. This simulation extends a fundamental
mechanism of eruption initiation as demonstrated in Paper A to failed eruption.
In the model, magnetic reconnection
has fundamental importance for formation of a pre-eruption MFR, initiation of its
eruption and cause of its failure in producing a CME. The MFR is
created by slow tether-cutting reconnection at the current sheet that
is formed in the core field as driven by photospheric shearing and
converging flows. The eruption, as indicated by fast rise of the MFR,
is caused by reconnection that becomes progressively faster in the
core field.  The failure and broken up of the erupting MFR is also
owing to reconnection, but is an external type which occurs in the
interface between the MFR and the overlying field. Our model can
explain well the observations of~\citet{zhouWhyTorusunstableSolar2019}
that some failed filament eruptions, although with the erupting filament reaching well above the
torus-unstable heights, often has a large rotation during its rise. This is because the large rotation of the MFR (i.e., the
filament) can result in large angle between the MFR field lines and
the overlying ones. Therefore an external current sheet can be formed
between the upward rising MFR and the overlying flux and reconnection
occurs in this current sheet can ruin the MFR. Such an external reconnection
also occurred in the simulations
of~\citet{hassaninhelicalkinkinstability2016} and \citet{chenModelConfinedSolar2023}, in which an ideally unstable
MFR (of either kink instability or torus instability) erupts and interacts with the background field through
reconnection.
In future works, we will investigate what are the key conditions in
determining the failure of the eruption, with numerical experiments by
using background fields in the parameter space of different strengths,
different flux distributions, different flux ratios as well as
different angles between them and the core field.

\section*{Acknowledgements}
This work is jointly supported by Shenzhen Science and Technology
Program (Grant No. RCJC20210609104422048), Shenzhen Technology Project
JCYJ20190806142609035, Shenzhen Key Laboratory Launching Project
(No. ZDSYS20210702140800001), Guangdong Basic and Applied Basic
Research Foundation (2023B1515040021) and National Natural Science
Foundation of China (NSFC 42174200). The computational work was
carried out on TianHe-1(A), National Supercomputer Center in Tianjin,
China. 

\section*{Data Availability}
All the data generated for this paper are available from the authors
upon request.



\input{manu_failederuption.bbl}








\bsp	
\label{lastpage}
\end{document}